\documentclass[aps,prr,reprint]{revtex4-2}
\usepackage{blindtext}
\usepackage[section]{placeins}
\usepackage[dvipsnames]{xcolor}
\usepackage{xfrac}
\usepackage{amsmath}
\usepackage{amsfonts}
\usepackage{amssymb}

 \newcounter{thm}
 \newcounter{ex}
 \newcounter{re}
\let\oldequation\equation
\let\oldendequation\endequation
\providecommand{\keywords}[1]
{
  \small	
  \textbf{\textit{Keywords---}} #1
}
\renewenvironment{equation}
  {\linenomathNonumbers\oldequation}
  {\oldendequation\endlinenomath}
\usepackage{graphicx}
\usepackage{dcolumn}
\usepackage{bm}
\usepackage{amsmath,verbatim}
\usepackage{flushend,balance}

\usepackage[mathlines]{lineno}
\modulolinenumbers[5]
\linenumbers\relax 
\begin{document}
\nolinenumbers
\preprint{AAPM/123-QED}

 \title{Low probability states, data statistics, and entropy estimation}

\author{Dami\'{a}n G. Hern\'{a}ndez $^{1,4,*}$}
\author{Ahmed Roman $^{1,*}$}
\author{Ilya Nemenman$^{1,2,3}$} 
\affiliation{$^{1}$Physics Department, Emory University, Atlanta, Georgia, USA\\
$^{2}$ Biology Department, Emory University, Atlanta, Georgia, USA\\
$^{3}$ Initiative for Theory and Modeling of Living Systems, Emory University, Atlanta, Georgia, USA\\
$^{4}$ Department of Medical Physics, Centro Atómico Bariloche and Instituto Balseiro, 8400 San Carlos de Bariloche, Argentina\\
$^*$ D.G.H. and A.R. contributed equally to this work.}

\date{\today}

\begin{abstract}
A fundamental problem in analysis of complex systems is getting a reliable estimate of entropy of their probability distributions over the state space. This is difficult because unsampled states can contribute substantially to the entropy, while they do not contribute to the Maximum Likelihood estimator of entropy, which replaces probabilities by the observed frequencies. Bayesian estimators overcome this obstacle by introducing a model of the low-probability tail of the probability distribution. Which statistical features of the observed data determine the model of the tail, and hence the output of such estimators, remains unclear.   Here we show that well-known entropy estimators for probability distributions on discrete state spaces model the structure of the low probability tail based largely on few statistics of the data: the sample size, the Maximum Likelihood estimate, the number of coincidences among the samples,  the dispersion of the coincidences. We derive approximate analytical entropy estimators for undersampled distributions based on these statistics,  and we use the results to propose an intuitive understanding of how the Bayesian entropy estimators work.
\end{abstract}

\maketitle


\section{Introduction} Estimating entropy -- that is, the measure of uncertainty \cite{shannon1948mathematical,cover2012elements} -- of a random variable from its samples is often a key question in analysis of complex systems. This estimation from a finite (and often small) set of samples is a hard problem, especially for high dimensional systems, where the number of states that a variable can take quickly overwhelms the number of samples $N$. Then many of the states, hereafter called {\em low probability states}, have probability $<1/N$. Collectively, we refer to all of these states as the {\em tail} of the probability distribution. While there may be a lot of samples in the tail, each low probability state will not be sampled typically, or will be sampled at most once.  Because of the tail, the entropy estimator that replaces  probabilities of states by their empirical frequencies  (the so called {\em naive} or {\em Maximum Likelihood} estimator \cite{strong1998entropy}) has a large sample size dependent bias \cite{paninski2003estimation}.  Corrections have been derived to overcome this bias \cite{miller1955note, grassberger2003entropy, berry2013simple}, but these  tend to be valid only in the well-sampled regime. Outside of this regime, Bayesian  \cite{wolpert1995estimating, nemenman2002entropy, archer2014bayesian} and some non-parametric \cite{chao2003nonparametric, chao2013entropy, cerquetti2019exact} estimators may still result in  low bias estimates by imposing {\em a priori} assumptions on the probabilities of the low-probability states.

Although these Bayesian and non-parametric estimators perform well on some data sets, it is known that no estimator can be universally unbiased in this regime \cite{paninski2003estimation,Antos01}. Thus it is crucial to understand how these estimators extract information about entropy from  data, and hence when they will fail. Unfortunately, such theoretical understanding is missing for many estimators. Ma was the first to point out that estimation of entropy is possible for poorly-sampled uniform  distributions by analysing a particular statistics of the data: {\em coincidences} \cite{ma1981calculation}. Nemenman extended the theoretical idea that coincidences determine entropy to non-uniform distributions obeying some Bayesian priors  \cite{nemenman2011coincidences}. However, a similar theoretical understanding  is still missing in a broader context, and it remains unclear which statistics of data, in addition to the number of coincidences, may contribute to entropy estimation and why.

In this paper, we analytically investigate two Bayesian estimators: that of Nemenman, Shafee and Bialek \cite{nemenman2002entropy,nemenman2004entropy} and  of Archer and Pillow \cite{archer2014bayesian}.  We focus on the regime, which is arguably the most important for real life applications, where the number of states with at least one sample,  $K_1$, is similar to the total number of samples, $K_1 \sim N \gg 1$, and yet $K_1<N$, so that there are coincidences in the data. Outside of this regime, the probability distribution is either well-sampled (so that many different methods for entropy estimation would work), or there are  no coincidences at all (so that entropy estimation is impossible). In our regime of interest, we show that the result of the estimation by the studied estimators depends on the Maximum Likelihood entropy estimate $S_0$, the number of  coincidences, and also on two measures of  {\em dispersion of coincidences}.  The first of these, $K_2$, is the number of states with at least two samples. The second, which we call $Q_1$, characterizes the spread of coincidences over states with three or more samples. 

\begin{figure*}[!t]
\centering 
\includegraphics[width=\textwidth]{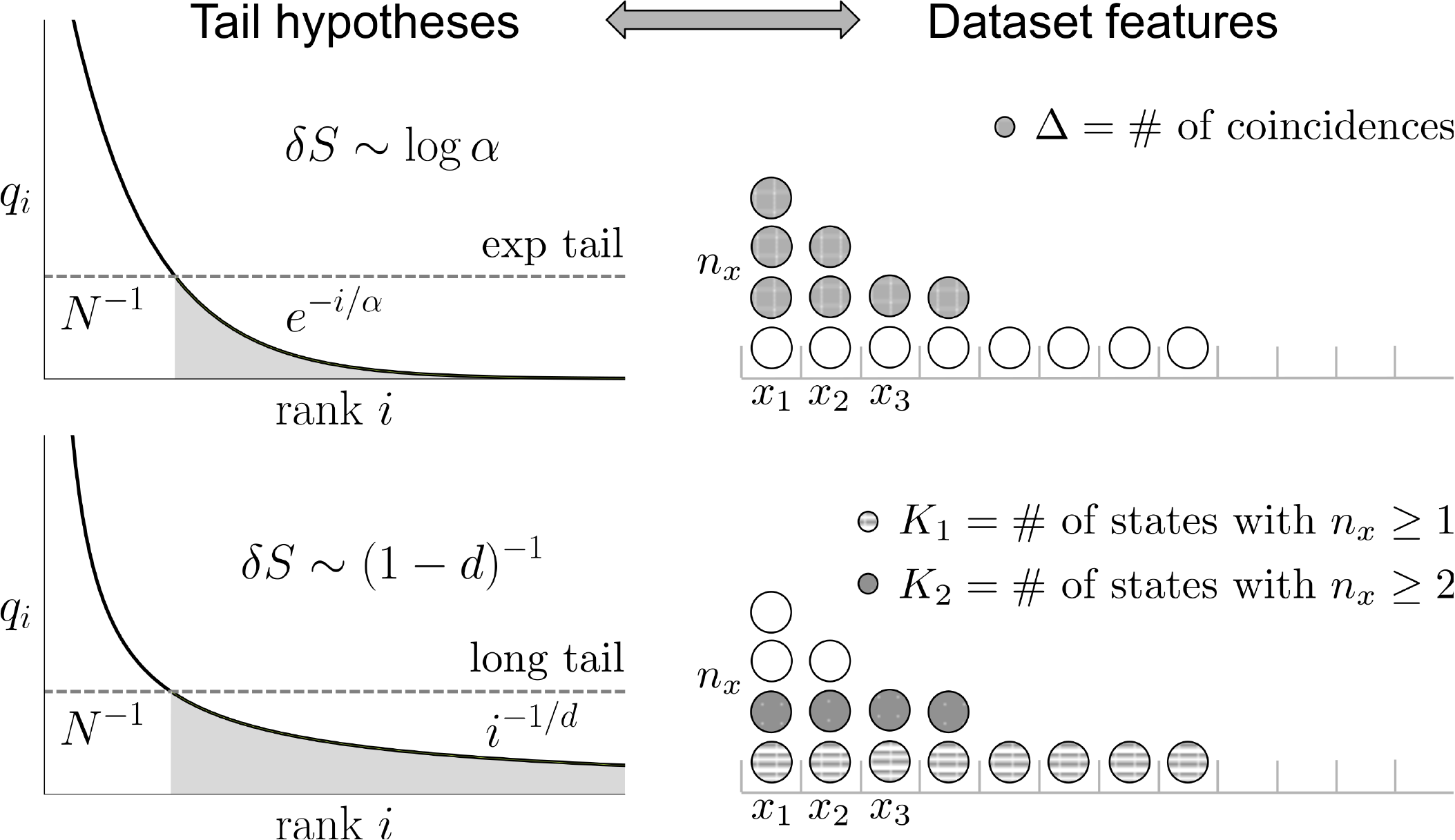}
\caption{Relation between assumptions about the tail structure and the statistics that determine entropy estimation. The set of unsampled states, $q_i \leq 1/N$, which we refer to as the {\em tail}, may contribute substantially to the entropy. However,  the Maximum Likelihood estimation overlooks this contribution. If the rank ordered plot of the tail is exponential with the scale $\alpha$ (top panel), then the tail has effectively $\alpha$ states, which contribute $\delta S\sim \log\alpha$ to the entropy. While the tail cannot be observed directly, it pulls samples from the head of the distribution, so that the number of coincidences, $\Delta$, in the head decreases as $\alpha$ grows. Thus one can estimate $\alpha$ and hence the entropy itself from $\Delta$.  Alternatively, if the rank-ordered plot of the tail has a power law structure with the exponent $-1/d$, then the tail does not have a finite effective size (bottom panels). Then its contribution to entropy depends on $d$ as $\delta S \sim (1-d)^{-1}$. In this case, one can estimate $d$, and hence the entropy, from the dispersion of the coincidences, which depends, in part, on how many samples happen once or more, $K_1$, or twice or more, $K_2$, in the dataset.}
\label{fig1}
\end{figure*}

We show that values of these statistics are related to the structure of the tails of the probability distribution that is assumed by the estimators. Specifically, a short, exponential, tail is more likely to be inferred by the estimators when there many coincidences or they are dispersed. If the number of coincidences is intermediate, and the coincidences are concentrated, then the estimators infer a long tail. In between these two regions, a mixed tail dominates. We show that the studied estimators correct Maximum Likelihood, and that the correction is larger when there are fewer coincidences and they are concentrated, which in turn happens with a large exponential tail or a slowly-decaying long tail. This understanding relates the observable data statistics to assumptions that Bayesian estimators make about the underlying probability distributions (see Fig.~\ref{fig1}), and hence provides an intuitive explanation for how these estimators work and, crucially, when they fail.

\section{Overview of Bayesian entropy estimation}

Given a probability distribution $\{q_x\}=\bm{q}$ for a discrete one-dimensional random variable $X$, its entropy is defined as \cite{shannon1948mathematical}
\begin{equation}
S(\bm{q})=-\sum_{x} q_x \log q_x.
\label{s2e1}
\end{equation}
Note that we use the natural logarithm throughout this paper, and hence entropy is measured in {\em nats}. One is often faced with a problem when $S$ must be estimated for unknown $q_x$ from a set of $N$ samples $\{x_1, \dots, x_N\}$ from the probability distribution. The Maximum Likelihood estimator of entropy, $S_0$, is then defined by replacing the probabilities with frequencies $q_x\to \hat{q}_x= n_x/N$, 
\begin{equation}
S_0=S(\hat{\bm{q}})=-\sum_{x} \frac{n_x}{N} \log \frac{n_x}{N}. \label{MLEstimator}
\end{equation}
States with zero frequencies in the sample do not contribute to $S_0$ resulting typically in underestimation  of the entropy \cite{paninski2003estimation}.   In general, because of this low probability tail, estimation of entropy from data is very hard  when the number of samples is smaller than the number of effective states of the variable, $N \ll \exp(S)$. 

Bayesian estimators address the problem by imposing various {\em a priori} assumptions $p(\bm{q})$. One then uses Bayes theorem to infer the {\em a posteriori} distribution of $\bm{q}$, and finally integrates over $\bm{q}$ to get the {\em a posteriori} distribution or moments of entropy. Specifically, the mean posterior entropy $\hat{S}=\langle S|\bm{n}\rangle$ given the counts $\bm{n}=\{n_x\}$ of how many times state $x$ was sampled is given by
\begin{align}
\hat{S}&=\langle S|\bm{n}\rangle = \int S(\bm{q})p(S|\bm{q})p(\bm{q}|\bm{n})d\bm{q} \nonumber \\ 
&= \int S(\bm{q}) \delta\left(S+\sum_{x}q_x \log q_x\right) p(\bm{q}|\bm{n})d\bm{q},
\end{align}
where $p(\bm{q}|\bm{n})$ is the posterior over $\bm{q}$ under some prior $p(\bm{q})$,
\begin{equation}
   p(\bm{q}|\bm{n})=\frac{p(\bm{n}|\bm{q})p(\bm{q})}{p(\bm{n})} = \frac{\prod_x q_x^{n_x}p(\bm{q})}{p(\bm{n})}.
   \label{eq:posterior}
\end{equation}
For distributions with known finite size $\mathcal{A}$ of the space of the possible outcomes (aka the {\em alphabet size}), the Dirichlet distribution  is often chosen as a prior due to its conjugacy with the categorical distribution: 
\begin{equation}
   p(\bm{q})= \text{Dirichlet}(\bm{q}|\lambda) \propto \prod_{i=1}^{\mathcal{A}}q_i^\lambda,
\end{equation}
where $\lambda$ is known as the concentration parameter. 

Note that any chosen  prior $p(\bm{q})$ implicitly imposes assumptions on the structure of the low probability tail (and hence its contribution to the entropy) based on the observed statistics of the well-sampled part of the probability distribution. However, these implicit assumptions usually are not made explicit, and they remain mysterious even for most commonly used Bayesian estimators. Lifting this veil is the goal of this work.

\subsection{The Nemenman-Shafee-Bialek (NSB) Estimator}
Nemenman et al.\ \cite{nemenman2002entropy} showed that, for variables with the finite alphabet size $\mathcal{A}$, Dirichlet priors on $\bm{q}$ with a fixed  value for the concentration parameter $\lambda$ correspond to highly concentrated \textit{a priori} distribution on entropy, which persists for large sample sizes. This bias induces incorrect entropy estimates, which nonetheless have low variance and hence are certain about their outputs. To address this issue, Ref.~\cite{nemenman2002entropy} suggested a Dirichlet-mixture prior
\begin{equation}
p_{\rm NSB}(\bm{q}) = \int \text{Dirichlet}(q|\lambda)p_{\rm prior}(\lambda)d\lambda,    
\end{equation} where $p(\lambda)$ are the mixture weights determined by 
\begin{equation}
p_{\rm prior}(\lambda) \propto \partial_\lambda \langle S|\lambda\rangle= \mathcal{A} \psi_1(\mathcal{A}\lambda+1)-\psi_1(\lambda+1),
\end{equation}
and where $\langle S|\lambda\rangle$ is the {\em a priori} expected entropy under the $\text{Dirichlet}(\bm{q}|\lambda)$ prior, and $\psi_1(\cdot)$ is the tri-gamma function \cite{abramowitz+stegun}. This choice of weights implies a nearly uniform {\em a priori} distribution for the entropy $S$ on the interval $[0,\log \mathcal{A}].$ The resulting entropy estimate is then 
\begin{align}
    \hat{S}_{\text{NSB}} &= \langle S|\bm{n} \rangle = \int \int S(\bm{q})p(\bm{q}|\bm{n},\lambda)p(\lambda|\bm{n})d\bm{q}d\lambda \nonumber \\
    &= \int  \langle S|\bm{n},\lambda\rangle \frac{p(\bm{n}|\lambda)p_{\rm prior}(\lambda)}{p(\bm{n})}d\lambda.
\end{align}
Here $ \langle S|\bm{n},\lambda\rangle$ is the posterior mean entropy under the prior $\text{Dirichlet}(\bm{q}|\lambda)$, and $p(\bm{n}|\lambda)$ is the evidence (which has a Polya distribution) \cite{minka00},
\begin{align}
    p(\bm{n}|\lambda) &= \int p(\bm{n}|\bm{q})p(\bm{q}|\lambda)d\bm{q} \nonumber \\
    &= \frac{N! \Gamma(\mathcal{A}\lambda)}{\Gamma(\lambda)^{\mathcal{A}}\Gamma(N+\mathcal{A}\lambda)} \prod_{i=1}^{\mathcal{A}} \frac{\Gamma(n_i+\lambda)}{n_i!} 
\end{align}
where $\Gamma(\cdot)$ is the gamma function \cite{abramowitz+stegun}.
Using the analytical expressions for the first two moments of posterior mean entropy  $ \langle S|\bm{n},\lambda\rangle$ (available from Refs.~\cite{wolpert1995estimating,nemenman2002entropy}), one then uses one-dimensional numerical integration over $\lambda$ to obtain  $\hat{S}_{\text{NSB}}$.

\subsection{The Dirichlet and the Pitman-Yor Processes}
\label{DP-PYP}
When the size of the state space is unknown or infinite, the standard NSB construction does not work. Then one commonly uses one of the following two stochastic processes to construct a prior $p(\bm{q})$ over a countably infinite state space: the Pitman-Yor Process (PYP) \cite{pitman1997two} and its special case, the Dirichlet Process (DP) \cite{Ferguson73}. To specify these processes, one requires two inputs: a parameter vector and a base distribution. Parameters of the Pitman-Yor process are known as the discount parameter $d$, $0\le d < 1$, and the concentration parameter $\alpha$. The parameters control the shape of typical distributions generated by the process. Specifically, $d$ controls the structure  of the low probability tail of $\bm{q}$, so that the tail typically decays as $q_x\propto x^{-1/d}$. The concentration parameter $\alpha$ control the probability mass  near the head of the distribution.  In the limit $d\rightarrow 0$,   $\text{PYP}(d,\alpha)$ becomes the Dirichlet Process, $\text{DP}(\alpha)$. In other words, the Dirichlet Process generates distributions with short tails. 

When the base distribution is the Beta distribution, one draws samples $q_x\sim \text{PYP}(d,\alpha)$ via the so called  {\em stick-breaking process} \cite{Ishwaran01}, which  uses an infinite sequence of independent Beta-distributed random variables $\beta_x \sim \text{Beta}(1-d,\alpha+ x d)$, so that
\begin{equation}
    \Tilde{q}_x = \beta_x\prod_{y=1}^{x-1}(1-\beta_y).
\end{equation}
Thus obtained $\Tilde{\bm{q}}$ are not strictly decreasing with $x$, and so one obtains a strictly non-increasing distribution $\bm{q}$ from them by rank ordering. 

\subsection{Expectations over DP and PYP Posteriors}
Previous studies \cite{Ishwaran03} showed that PYP priors (for multinomial observations) yield a posterior $p(\bm{q}|\bm{n},\alpha , d)$, which consists of two parts: probability of $K_1$ states that exist in the sample with the counts of, at least, one, and probability of states that are not sampled. We will denote the set of states with nonzero counts as $\mathbb{K}$, and its cardinality is $K_1=||\mathbb{K}||$. Then the first term of the posterior is given by the  Dirichlet distribution, $p(\bm{q}\in\mathbb{K}|\bm{\mu})\propto \prod_x q_x^{\mu_x}$, where $\bm{\mu}$ is a concentration vector $\bm{\mu}=(n_1-d,\cdots,n_{K_1}-d,\alpha+K_1 d)$. This leaves the probability of $q_* =1-\sum_{x\in\mathbb{K}}q_x$ for the unobserved states. In other words, the states with nonzero counts contribute the following to the posterior:
\begin{align}
  p(\bm{q}\in \mathbb{K}|\bm{n})&=p(q_1,\cdots,q_{K_1},q_*|\bm{n}) \nonumber \\ 
  &= \text{Dirichlet}(n_1-d,\cdots,n_{K_1}-d,\alpha+K_1 d) \nonumber \\
  &\propto q_*^{\alpha+K_1d}\prod_{i=1}^{K_1} q_i^{n_i-d}. \label{DirDistribution}
\end{align}
For the states that have no samples, the posterior is equal to the prior. Thus their contribution to the posterior is the Pitman-Yor Process, normalized by their total probability being $q^*$:
\begin{equation}
 p(\bm{q}\not\in \mathbb{K})= p(q_{K_1+1},q_{K_1+2},\cdots)= q^* \text{PYP}(d,\alpha+K_1 d).    
\end{equation}
Overall, this yields a closed form solution for the posterior mean and variance of the entropy $S$. Specifically, the resulting posterior mean $\langle S|\bm{n},\alpha , d\rangle$ is     
\begin{equation}
\begin{split}
\langle S|\bm{n},\alpha , d\rangle = \psi(\alpha+N+1)-\frac{\alpha+K_1d}{\alpha+N}\psi(1-d) \\
-\frac{1}{\alpha+N}\left(\sum_{x=1}^{K_1} (n_x-d)\psi(n_x-d+1)\right),
\end{split}
\label{PYPEntropy}
\end{equation}
where  $\psi(x)=\partial_{x}\log\Gamma(x)$ is the di-gamma function \cite{abramowitz+stegun}. Unfortunately, this is usually not a good estimate of entropy since, for fixed $\alpha$ and $d$,  the prior $\text{PYP}(d, \alpha)$ on $\bm{q}$ corresponds to a highly concentrated {\em a priori} distribution on entropy, just like was noted before in the context of the NSB estimator. To counter this, Archer and Pillow \cite{archer2014bayesian} followed the NSB prescription  and introduced a prior (mixture) over the parameters of $PYP(d,\alpha )$,  $p_{\rm prior}(\alpha , d)$,  which  uniformized the induced prior over entropy (with the caveat that, for a distribution on a countable alphabet, the entropy may be infinite, and hence strict uniform distribution over entropy is impossible). Specifically, they used
\begin{align}
&p_{\rm prior}(\alpha , d) = p(\gamma) = e^{-10/(1-\gamma)}, \quad \quad {\rm where}\\
&\gamma= (\psi(1)-\psi(1-d))/(\psi(\alpha+1)-\psi(1-d)) \label{gamma},
\end{align}
and then they confirmed numerically that this choice of the prior leads to good estimates of entropy for various test data sets.  In other words, they proposed a  new estimate of entropy, the Pitman-Yor Mixture (PYM): 
\begin{align}
\hat{S}_{PYM} &= \langle S|\bm{n}\rangle =  \int \langle S|\bm{n},\alpha , d\rangle p_{\rm posterior}(\alpha , d|\bm{n})d(\alpha , d) \nonumber \\
&= \int \langle S|\bm{n},\alpha , d\rangle \frac{p(\bm{n}|\alpha , d)p_{\rm prior}(\alpha , d)}{p(\bm{n})}d(\alpha , d),
\label{estimator}
\end{align}
where $\langle S|\bm{n},\alpha , d\rangle$ is given in Eq.~(\ref{PYPEntropy}). The evidence $p(\bm{n}|\alpha , d)$ is then given by  (see Ref.~\cite{archer2014bayesian} for a detailed derivation)
\begin{equation}
p(\bm{n}|\alpha , d) = \frac{\Gamma(1+\alpha)\prod_{l=1}^{K_1} (\alpha+l d)\prod_{x=1}^{K_1}\Gamma(n_x-d)}{\Gamma(1-d)^{K_1} \Gamma(\alpha+N)}.   
\label{evidence}
\end{equation}

Note that taking $d\to 0$ in Eqs.~(\ref{estimator} and \ref{evidence}) and making the identification $\alpha = \mathcal{A}\lambda$ in the limits $\lambda\rightarrow 0$ and $\mathcal{A}\rightarrow \infty$ such that $\alpha$ is finite, result in a countably-infinite analogue of the NSB estimator.

\section{Determining data statistics that define  entropy estimates}
In the section, we approximate the likelihood function of the Pitman-Yor process,  Eq.~(\ref{evidence}), analytically in terms of coincidence-based data statistics. We then numerically show that the resulting analytical entropy estimates are close to the exact Pitman-Yor Mixture estimator. We focus on  the regime where the Maximum Likelihood entropy estimator fails dramatically. For this, we study random variables with many accessible states in the regime where the number of unique samples, $K_1$, is of the order of the total sample size $N$. This regime corresponds to $K_1 \lesssim N\le\exp(S)$, where $N$ is the number of samples and $S$ is the true entropy.  


We start by considering the log-likelihood function, which is the logarithm of the evidence $p(\bm{n}|\alpha , d)$ in Eq.~(\ref{evidence}):
\begin{multline}
\mathcal{L}({\bf n}|\alpha , d)=\log\Gamma(1+\alpha)-\log\Gamma(N+\alpha)+\log\Gamma\left(\frac{\alpha}{d}+K_1\right)\\
-\log\Gamma\left(\frac{\alpha}{d}+1\right)
+\sum_{i=1}^{K_1}\log\Gamma(n_i-d)-K_1 \log\Gamma(1-d).
\label{logLikelihood}
\end{multline}
We now define $K_m$ as the number of states with at least $m$ counts in the total sample of size $N$, $K_m=\sum_{n_i \ge m}1$. We denote by $m_f$ the largest occupancy of any state in the sample. Further, we define  $\mathcal{K}$ as the vector, whose $m$th element is $K_m$.  We note that, for any function $f(n)$, 
\begin{equation}
     \sum_i f(n_i) = \sum_m (K_m-K_{m+1}) f(m).
     \label{identity}
\end{equation}
Thus, in particular, the log-likelihood $\mathcal{L}({\bf n}|\alpha , d)$ can be viewed as $\mathcal{L}({\mathcal{K}}|\alpha , d)$. 
With this, we can expand Eq.~(\ref{logLikelihood}) around $d=0$ to get (see Appendix \ref{A1:approxL} for details):
\begin{multline}
\mathcal{L}(\bm{n}|\alpha , d)\approx \mathcal{L}_a(\mathcal{K}|\alpha , d)\equiv \log\Gamma(1+\alpha)\\-\log\Gamma(N+\alpha)+\log\Gamma\left(\frac{\alpha}{d}+K_1\right)-\log\Gamma\left(\frac{\alpha}{d}+1\right)\\
+(K_1-1)\log d+K_2\log(1-d)-Q_1 d+\mathcal{O}(d^2),
\label{eq02:L}
\end{multline}
where 
\begin{equation}
    Q_1=\sum_{m=3}^{m_f}\frac{K_m}{m-1},
\end{equation}
and the subscript $a$ denotes the $d\to0$ asymptotic nature of the expression.

\begin{figure*}[!t]
\includegraphics[width=\textwidth]{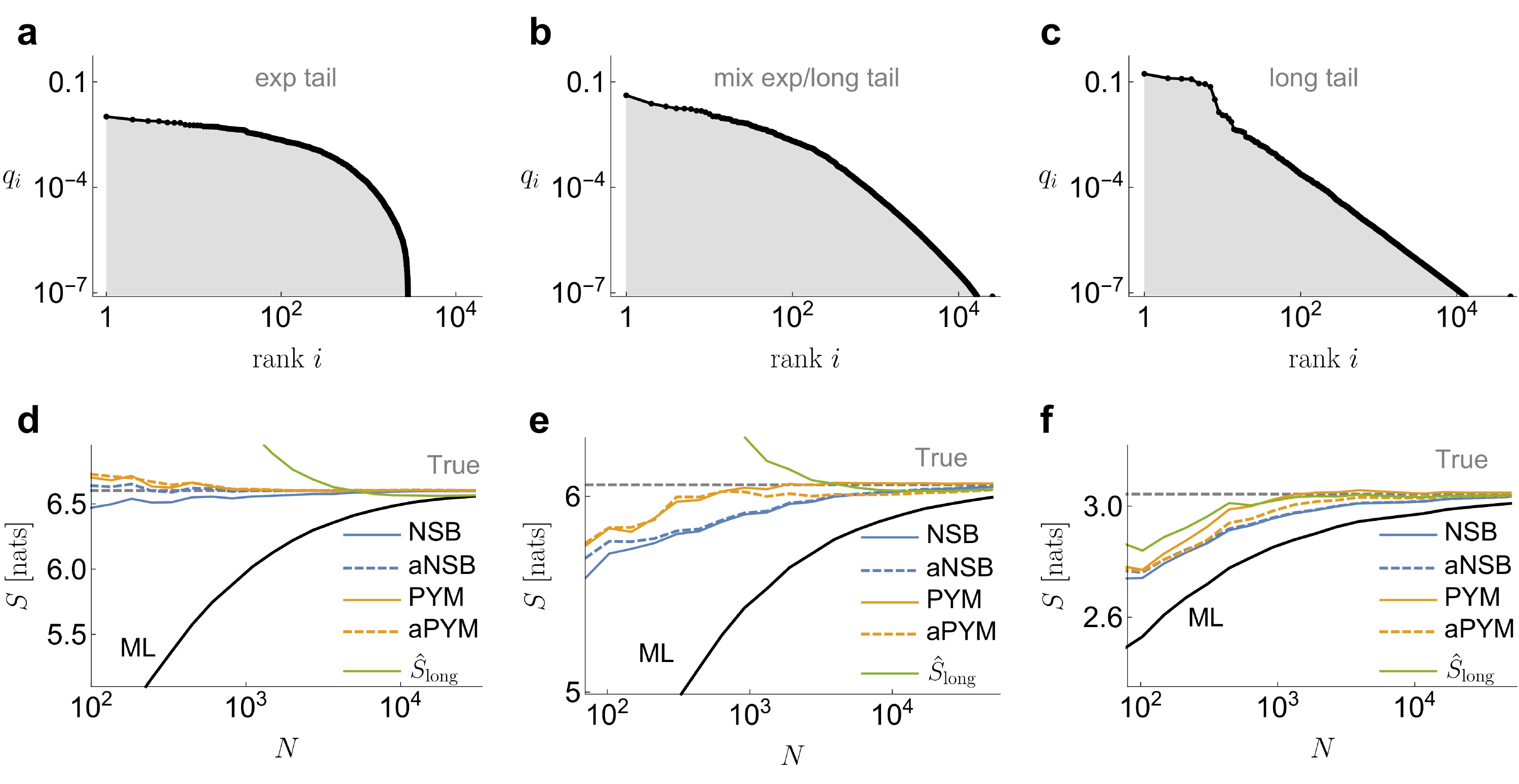}
\caption{Comparison between PYM and related estimators and their approximations for distributions with different tails. The upper panels ({\bf a}-{\bf c}) show the distributions, whose entropy is being estimated. The lower panels ({\bf d}-{\bf f})  show the corresponding  entropy estimates as a function of the number of samples, averaged over ten sets of samples. The full estimators, PYM and NSB (with a large alphabet size $\mathcal{A} = 20K_1$), almost overlap with our approximations, aPYM and aNSB. In all panels, we show results for Maximum Likelihood (black), NSB (blue), aNSB (dashed blue), PYM (orange), aPYM (dashed orange), and $\hat{S}_{\text{long}}$ (green) estimators. The dashed gray line represents the true value of entropy for each of the studied distributions.}
\label{fig2}
\end{figure*}

By rewriting the Maximum Likelihood estimate $S_0$ of Eq.~(\ref{MLEstimator}) in terms of coincidences (see Appendix~\ref{A3:approxS}),  using the identity Eq.~(\ref{identity}), and approximating certain terms that are finite in the limit $d\rightarrow 1$ via a Taylor expansion around $d\ll 1$, the mean posterior entropy, Eq.~(\ref{PYPEntropy}), results in (see Appendix~\ref{A2:approxS}):
\begin{multline}
\langle S|{\bf n},\alpha , d \rangle \approx
\langle S|{\mathcal{K}}, \alpha , d \rangle_a \equiv \psi(N+\alpha+1)\\-
 \left(\frac{\alpha+K_1}{\alpha+N}\right)\psi(1-d)
  +\frac{1}{\alpha+N}\Bigg[ .N(S_{0}-\log N)-K_1 \\ +K_2(\log 4-1-\psi(2-d))+Q_1 d \\ +\mathcal{O}\left(d^2,\sum_{m=3} \frac{K_m}{(m-1)^2}\right) \Bigg],
    \label{eq03:S}
\end{multline}
where $\mathcal{O}(d^2,\sum_{m=3}K_m/m^2)$ means that we kept terms that are at most linear in $d$ and at most proportional to $\sum_{m=3} \frac{K_m}{(m-1)}$. Interestingly, within this approximation, the log-likelihood and the posterior mean entropy depend on the sample size $N$, the Maximum Likelihood entropy estimate $S_0$, and the three characteristics of the coincidence vector: $K_1,K_2$ and $Q_1$.

The final step in approximating  the estimator $\hat{S}_{PYM}$, Eq.~(\ref{estimator}), is to integrate the expected entropy for fixed hyper-parameters $\langle S|{\mathcal{K}},\alpha,d\rangle_a$ over the posterior $p_{\rm posterior}(\alpha,d|{\bf n})\propto p({\bf n}|\alpha,d)p_{\rm prior}(\alpha,d)$ to form the Pitman-Yor mixture. Then the variance of the resulting estimator is dominated by the contribution from the uncertainty in the  posterior distribution of the parameters $\alpha,d$, which is about $80\%$ of the total variance in our simulations. 

This procedure of replacing $\langle S|\bm{n},\alpha,d\rangle$ with the asymptotic expression $\langle S|{\mathcal{K}},\alpha,d\rangle_a$ in Eq.~(\ref{estimator}) leads to a new estimator of entropy, which we call {\em approximate PYM} estimator, or aPYM. This estimator is fully determined by just few data statistics, $N$, $S_0$, $K_1$, $K_2$, and $Q_1$. There are also two limiting cases of this estimator. First, by taking $d\to0$ in Eqs.~(\ref{eq02:L}, \ref{eq03:S}), we define the approximate version of the NSB limit of the PYM estimator on a countably infinite number of possible outcomes, which we denote as aNSB. At the other extreme, taking $\alpha\to0$  in Eqs.~(\ref{eq02:L}, \ref{eq03:S}), corresponds to a prior that favors distributions with long tails. We denote the corresponding estimator as  $\hat{S}_{\text{long}}$. 
\begin{figure*}[!t]
\includegraphics[width=\textwidth]{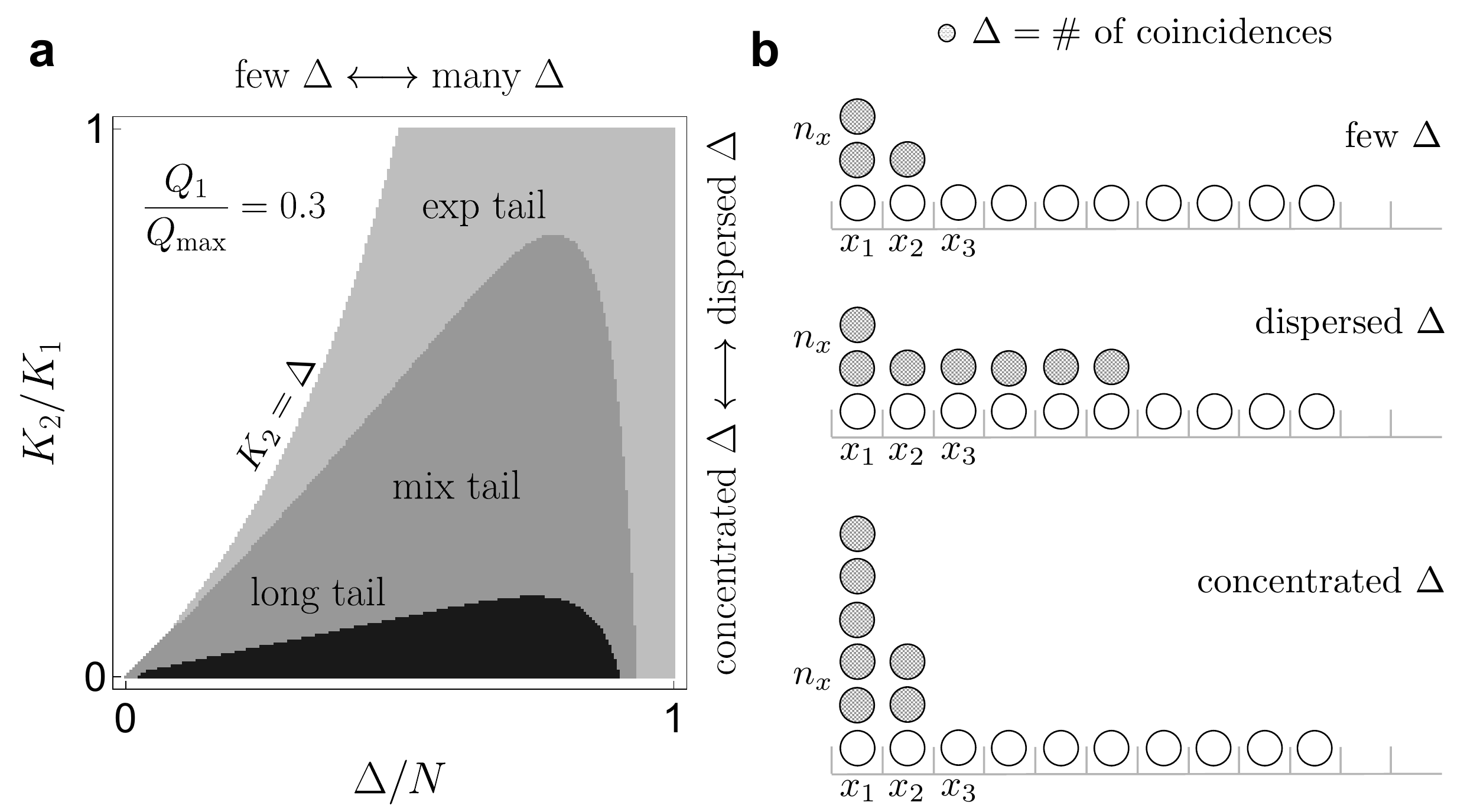}
\caption{{\bf a}: Phase diagram of the dominant tail hypothesis selected by the PYM estimator as a function of various statistics of the data sample. The explored statistics are the fraction of coincidences in the sample, $\Delta/N$, and dispersion of the coincidences,  $K_2/K_1$. This diagram is evaluated at the third crucial data  statistics set at $Q_1=0.3\,Q_{\text{max}}= 0.3 (\Delta-K_2)/2$.  {\bf b}: Schematic diagram that illustrates how sample sets with different $\Delta$, $K_1$, and $K_2$ may look like.  An empty or gray circle above a state $x_i$ represent a single sample  for that state. Gray circles denote  coincidences. }
\label{fig3}
\end{figure*}

The above observation that, in the undersampled regime where $\exp(S/2)<N<\exp(S)$, the PYM entropy estimator and its relatives are determined approximately by just few statistics of the data, $\{N,S_0,K_1,K_2,Q_1\}$, is the main result of our paper. To corroborate this, we explore the quality of the approximation numerically for  different distributions $\bm{q}$.  Figure~\ref{fig2} presents results for three distributions with different structures of tails, generated from the Pitman-Yor Process: a distribution with an exponential tail (Fig.~\ref{fig2}{\bf a}, ${\rm PYP}(d=0,\alpha=400)={\rm DP}(400)$), one with a mixed tail (Fig.~\ref{fig2}{\bf b}: ${\rm PYP}(d=0.4,\alpha=100)$), and one with a long tail (Fig.~\ref{fig2}{\bf c}: ${\rm PYP}(d=0.6,\alpha=0)$). In the lower panels we show the results of estimating entropy for different dataset sizes using the ML estimator, the PYM estimator, the NSB estimator with a large alphabet size $\mathcal{A} = 20K_1$, and the three approximations: aPYM, aNSB, and $\hat{S}_{\text{long}}$. All results are averaged over ten sets of random samples. In all cases, the differences between NSB and aNSB on the one hand, and PYM and aPYM on the other are negligible, supporting the accuracy of the approximation. All four of these estimators produce high quality estimates for all sample sizes. Further,  we  also checked that the approximation of the posterior error of the estimators is close to that of the full versions (not shown). In contrast, $\hat{S}_{\text{long}}$ only performs well when the distribution has a long tail, and the Maximum Likelihood never works well.

\section{Tail-hypothesis and entropy estimation phase diagrams}
\label{phasediagram}

The above discussion shows that the PYM estimator and its relatives work by first estimating the most likely $\alpha$ and $d$ from the sampled data, and then using these estimated parameters to approximate the structure of the low probability tail (from short, to long) and hence of its contributions to the entropy. We further showed that, in the regime of interest,  the log-likelihood of $\alpha$ and $d$ is dominated by just few statistics: $N$, $S_0$, $K_1$, $K_2$, and $Q_1$. It is thus illustrative to understand, which combinations of these statistics select which hypothesis on the structure of the tail. Building the corresponding phase diagram of the selected tail structure as a function of the data statistics is the goal of this Section. 

We will consider three classes of tails: exponential ($d=0$ selected, denoted as hypothesis $H=1$), long tail ($\alpha=0$ selected, denoted as hypothesis $H=2$), and a mixed tails (arbitrary $\alpha$ and $d$, denoted as $H=3$). Our goal is then to evaluate which of the three tail hypotheses has a higher probability given the data. Long and short tail hypotheses  have one parameter each, while the mixed tail hypothesis has two parameters and contains the other two hypotheses as special cases. Thus when evaluating the log-likelihoods of each of the hypotheses, we must penalize them for having a different number of parameters, which we do using Bayesian Information Criterion \cite{Schwarz78}. To do this, we evaluate the likelihoods
\begin{equation}
\mathcal{L}_H = \log p({\mathcal{K}}|\hat \alpha,\hat d)+\log p_{\rm prior}(\hat \alpha,\hat d)-\frac{n_H}{2}\log N, 
\label{eq03:aL}
\end{equation}
where $\hat\alpha$ and $\hat d$ are the maximum likelihood values of the parameters within each hypothesis, and $n_H$ is the number of parameters for the hypothesis ($n_H=2$ for $H=3$, and $n_H=1$ otherwise). We remind the reader that, by construction, $\hat\alpha = 0$ for the long tail hypothesis, $H=2$, and $\hat d =0$ for the short tailed hypothesis, $H=1$. 

We determine the regions of the $N,S_0,K_1,K_2,Q_1$ space, where one of the three $\mathcal{L}_H$ dominates, and plot the slice of this phase diagram in Fig.~\ref{fig3}. Specifically, in the Figure, we vary the total number of {\em coincidences}, $\Delta=N-K_1$, and the number of {\em states with coincidences}, that is, the number of states with more than two counts, $K_2$. By sampling many distributions, we empirically observe  that the value $Q_1\sim 0.6 (\Delta-K_2)/2$ is when the rest of the $\Delta-K_2$ counts are uniformly dispersed, and $Q_1$ tends to zero when the rest of the counts are concentrated in a single state. Note that the maximum value $Q_1$ can take is $Q_{\text{max}} = \frac{\Delta-K_2}{2}$. For this reason, we choose the intermediate representative value $Q_1=0.3 Q_{\max}=0.3\frac{\Delta-K_2}{2}$.

To simplify the presentation, we plot the winning tail hypothesis as a function of $\Delta/N$ and $K_2/K_1$. Normalized in this way, the diagram is constrained to a square of size 1, as $0\le \Delta/N, K_2/K_1\le 1$. In addition, $K_2\le \Delta$, which means that the upper left corner is not accessible. The ratio $\Delta/N$ determines how common are the coincidences, and the ratio $K_2/K_1$ describes whether the coincidences in the data are concentrates in a few states, or dispersed over many states (see Figure~\ref{fig3}b).



Figure~\ref{fig3}{\bf a} show that the exponential tail hypothesis dominates when  there are  many coincidences, $\Delta/N\sim 1$, or when the coincidences are dispersed, that is $K_2/K_1 \sim 1$ or $K_2/\Delta \sim 1$. Both cases can be explained as corresponding to distributions that are relatively uniform on some fixed number of states, and have zero probability elsewhere. A long tail only dominates when the fraction of coincidences has an intermediate value, but the coincidences are highly concentrated, $K_2/K_1\ll 1$. In other words, in this case, there are dominant states, but a lot of samples still fall outside of them. For other values of $\Delta/N$ and $K_2/K_1$, the mixed tail hypothesis dominates.

Equipped with this picture of which tail hypothesis is selected by the PYM estimator as a function of data statistics, we now can calculate how the estimator corrects the ML entropy value $S_0$ for different data statistics. Integrating the mean posterior entropy $\langle S|\mathcal{K},\alpha,d \rangle_a$, Eq.~(\ref{eq03:S}), over our approximation of the posterior, $p_a (\alpha,d|{\mathcal{K}})$, which we obtain by exponentiating Eq.~(\ref{eq02:L}), we get the approximate PYM estimator $\hat{S}_{PYM,a}$. The Maximum Likelihood estimate $S_0$ enters linearly in the posterior mean entropy, Eq.~(\ref{eq03:S}). Thus we write \begin{equation}
    \langle S|\mathcal{K},\alpha,d \rangle_a=  b_{\alpha,d} \,S_0 +\delta S_{\alpha,d},
\end{equation}
where $b_{\alpha,d}$ and  $\delta S_{\alpha,d}$ can be read off from Eq.~(\ref{eq03:S}). Performing the integral over the approximate posterior,  this becomes:
\begin{equation}
    \hat{S}=\delta S+b \,S_0,
    \label{S-hat}
\end{equation}
where $\delta S$ and $b$ are averages of the corresponding $\alpha$- and $d$-dependent quantities. Thus {\em independent} of the Maximum Likelihood entropy value, within our approximation, the PYM estimator obtains the entropy estimate by decreasing the ML contribution from the well-sampled head of the distribution and adding an offset that comes from the low probability tail.  This is similar to so-called partition-based entropy estimators, \cite{nemenman2004entropy, chao2013entropy,Srivastava17,Nemenman15}, which divide the state space into sub-spaces, estimate entropy in each sub-space, and then add the estimates weighted by the probability of being in a corresponding sub-space. However, here this partitioning arises naturally from the Bayesian framework within our approximations.

\begin{figure*}[!t]
\includegraphics[width=\textwidth]{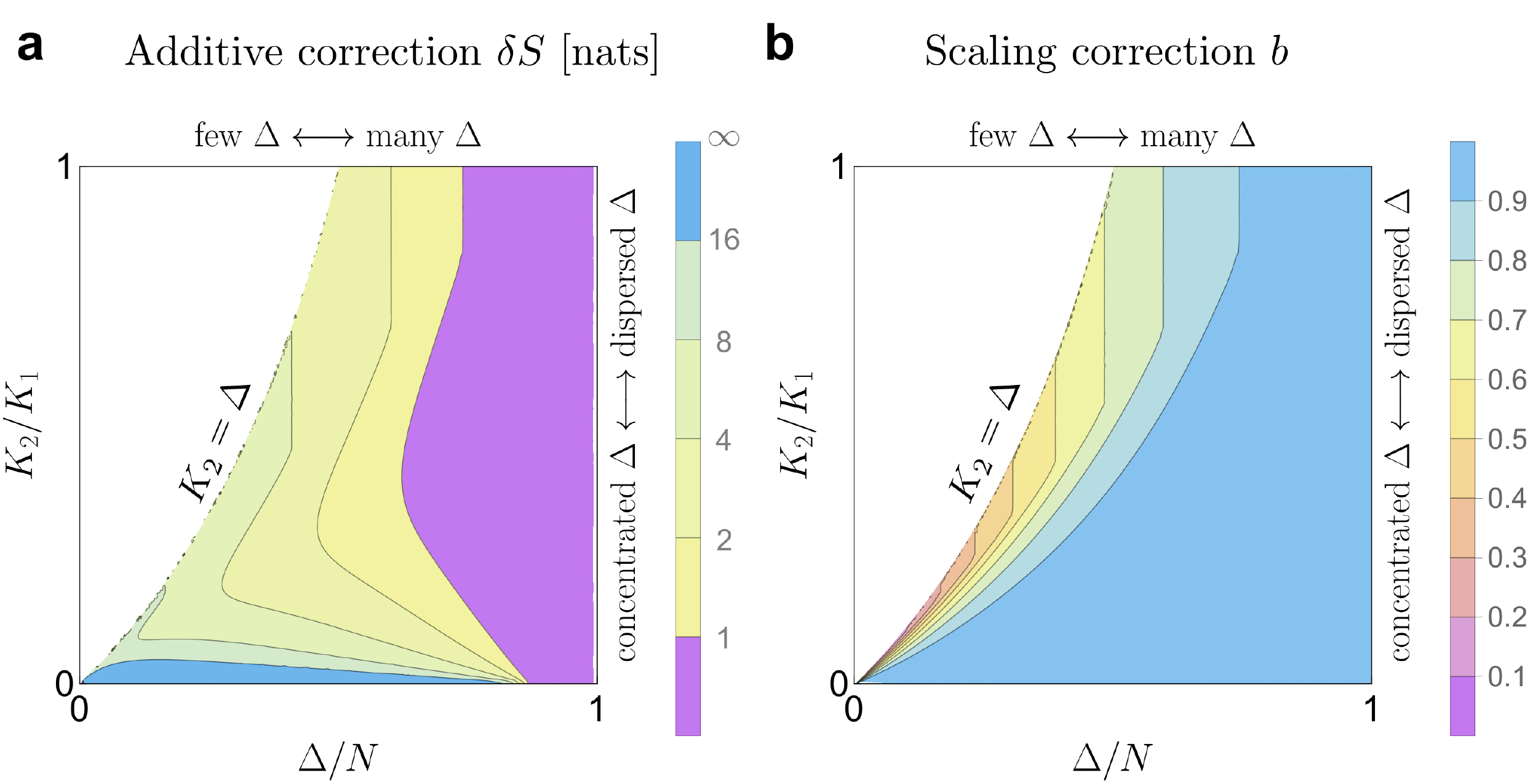}
\caption{Corrections to entropy estimation as a function of determining data statistics. We break down the final estimation for entropy in two parts, as $\hat{S}= \delta S(\sfrac{\Delta}{N},\sfrac{K_2}{K_1})+b(\sfrac{\Delta}{N},\sfrac{K_2}{K_1})\,S_0$, where $\delta S$ is the additive correction and $b$ is scaling factor or weight for the Maximum Likelihood estimate. Well-sampled distributions are located in the upper-right corner where $\delta S=0$ and $b=1$. As in the previous plots, we leave $Q_1=0.3\,(\Delta-K_2)/2$. {\bf a}: Additive correction to entropy. {\bf b}: Scaling correction to entropy.}
\label{fig4}
\end{figure*}

Both the scale factor and the offset depend on the dominant $\alpha$ and $d$ contributing to the estimator, and hence on the usual statistics of the data, $\Delta$, $K_1$, $K_2$, and $Q_1$. Specifically, we numerically observe that the value of $b$ obtained from Eq.~(\ref{S-hat}) satisfies 
\begin{equation}
b=\langle N/(\alpha+N)\rangle\le 1,
\end{equation}
where the average is over the product of the approximate posterior obtained by exponentiating Eq.~(\ref{eq02:L}) and the prior $p(\gamma) = e^{-7\gamma/100}$ with $\gamma$ defined in Eq.~\ref{gamma}. Note that $\alpha$ is a measure of how much probability is concentrated in the tail. Thus the ratio $N/(\alpha+N)$ approximates the overall weight of the the well-sampled head of the distribution, requiring to decrease the contribution to the entropy from the head by this factor. This matches our assertion that the aPYM estimator is a partition-based estimator, separating the head from the tail. 

In Figure~\ref{fig4} we show results of numerical estimation of the offset $\delta S$ and the scaling factor $b$ as a function of the fraction of coincidences, $\Delta/N$, and the dispersion of coincidences, $K_2/K_1$. As in the previous case, we keep $Q_1=0.3Q_{\rm max}$. We also set  $N=10^4$. Figure~\ref{fig4}(a) shows that the additive term grows when the fraction of coincidences $\Delta/N$ decreases, and when $K_2/K_1$ is small, so that coincidences are concentrated. Both of these cases correspond to a lot of mass in the tail (see corresponding long tail region in Figure~\ref{fig3}(a). The largest values of $\delta S$ occur along the boundary strip $(\Delta/N,K_2/K_1\ll 1)$ and the boundary $K_2=\Delta$. Panel {\bf b} shows that the scaling factor $b$ is close to 1 in most areas, except near the boundary edge $K_2=\Delta$. Along this boundary, the scaling factor becomes the largest when the number of coincidences decreases,  $\Delta/N\ll1$. Figure~\ref{fig4} clearly highlights when Bayesian corrections to the ML estimation of entropy are essential: regions of few and concentrated coincidences.

\section{Discussion}
The major finding of this work is an  excellent approximation for the PYM  estimator, one of the best Bayesian entropy estimators, and its various relatives (such as NSB). The approximation simplifies the numerics  considerably. Crucially, the approximation also shows that the output of the PYM entropy estimator depends on just a few statistics of the data, namely the maximum likelihood (ML) entropy estimate, the fraction of coincidences $\Delta/N$, and the dispersion of coincidences $K_1/K_2$, and $Q_1$. We showed that that workflow of the estimator can be interpreted as first estimating the parameters $d$ and $\alpha$ based on the aforementioned statistics, and with them the tail structure and the total weight of the tail. Then the estimator rescales the ML entropy estimate by the weight of the well-sampled head of the distribution, and adds to it the estimated entropy of the tail. The phase diagrams of which tail structure the estimator selects, Fig.~\ref{fig3}, and how it corrects the ML estimate, Fig.~\ref{fig4}, illustrate these points. 

Early work of Ma \cite{ma1981calculation} showed that when states are equiprobable, in the under-sampled regime, the coincidences in counts can help with the inference of the entropy of a system. Later Nemenman~\cite{nemenman2011coincidences} showed that in the severely under-sampled regime ($K_1$ close to $N$), entropy estimation depends on the number of coincidences $K_1$. Further, he pointed out how reliable entropy estimates may be obtained by partitioning the overall state space of the variable into sub-spaces with similar sampling properties \cite{Nemenman15}. Here we extend these results to the whole regime where entropy estimation is challenging for multinomial observations, $\exp(S/2)<N <\exp(S)$, by approximating the more general PYM estimator. Our identification of the small set of statistics, which define the output of the estimator, lifts the veil from its inner workings, allowing for a simple, semi-analytical estimation procedure. In particular, this allows us to predict if a particular estimator will be biased simply by looking at the values of the select statistics of the data.

How to match {\em a priori} assumptions about the underlying distributions to the data to allow for an unbiased estimation of quantities of interest---such as entropy \cite{nemenman2004entropy, archer2014bayesian} or the mutual information  \cite{hernandez2019estimating}--- is an open problem \cite{hernandez2022inferring}. It requires understanding the relation between the {\em a priori} assumptions and the data features that influence the inference. In this work, we build such a link for entropy estimation, and we hope that similar links might exist for other difficult estimation problems. 

\begin{acknowledgments}
This work was funded, in part, by the Simons Investigator award (IN), the Simons-Emory International Consortium  (AR and IN), NSF Grant 1822677 (DGH and IN), NIH Grant 2R01NS084844 (AR, IN), the International Physics of Living Systems Network (NSF Grant 1806833, AR). IN would like to acknowledge hospitality of the Aspen Center for Physics, funded in part by NSF Grant 1607611.
\end{acknowledgments}
\bibliography{PRE.bbl}

\section{Appendix}
\subsection{Marginal likelihood approximation for a Pitman-Yor process}\label{A1:approxL}
In this Appendix we show how to approximate the marginal posterior of a Pitman-Yor process in the regime $K_1\lesssim N\le \exp(S)$. We start by manipulating each term in the logarithm of the evidence $\mathcal{L}=\log p(\mathbf{n}|\alpha , d)$ from Eq.~(\ref{evidence}),
\begin{multline}
\mathcal{L}(\mathbf{n}|\alpha , d)=\sum_{l=1}^{K_1-1} \log(\alpha+l d)+\\\sum_{i=1}^{K_1} \log\Gamma(n_i-d)
-K_1 \log\Gamma(1-d)\\+\log\Gamma(1+\alpha)-\log\Gamma(N+\alpha).
\label{eq01}
\end{multline}

To simplify the first term in Eq.~(\ref{eq01}), we rewrite it in terms of coincidences $K_1$ as follows:
\begin{multline}
I_1   =\displaystyle\sum_{l=1}^{K_1-1} \log(\alpha+l d) \\= \sum_{l=1}^{K_1-1}\left[ \log d+ \log\left(\frac{\alpha}{d}+l\right) \right]
\displaystyle  = (K_1-1)\log d\\
+\sum_{l=1}^{K_1-1}\left[ \log\Gamma\left(\frac{\alpha}{d}+l+1\right)-\log\Gamma\left(\frac{\alpha}{d}+l\right)\right] \\
\displaystyle  = (K_1-1)\log d+\log\Gamma\left(\frac{\alpha}{d}+K_1\right)-\log\Gamma\left(\frac{\alpha}{d}+1\right).
\label{eq02}
\end{multline}
In order to rewrite the rest of the terms of Eq.~(\ref{eq01}) in terms of various coincidence statistics, we use the identity Eq.~(\ref{identity}).  Joining the second and third terms in Eq.~(\ref{eq01}) and rewriting them in terms of count multiplicities yields
\begin{align}
&\sum_{i=1}^{K_1} \log\Gamma(n_i-d)-K_1 \log\Gamma(1-d)  \nonumber\\
&= -K_2 \log\Gamma(1-d)+\sum_{m=2}(K_m-K_{m+1}) \log\Gamma(m-d) \nonumber\\
&=\sum_{m=2} K_m \left[ \log\Gamma(m-d)-\log\Gamma(m-1-d) \right] \nonumber\\
&=\sum_{m=2} K_m \log(m-1-d) \nonumber\\
&= K_2 \log(1-d)+Q(d),
\label{eq03}
\end{align}
where
\begin{equation}
Q(d)=\sum_{m=3}^{m_f} K_m \log(m-1-d).
\label{eq04}
\end{equation}
where $m_f$ denotes the largest occupancy of any state in the sample.
Since the domain of $0\le d <1$ is small, $Q(d)$ is approximately linearly varying with $d$, so that we can expand it around $d=0$:
\begin{multline}
 Q(d) = Q(0)- \sum_{j=1} \left[\sum_{m=3}\frac{K_m}{(m-1)^j}\right]\frac{d^j}{j} \nonumber\\
\approx Q(0)-\left[\sum_{m=3} \frac{K_m}{m-1} \right] d \\- \frac{1}{2}\left[\sum_{m=3} \frac{K_m}{(m-1)^2} \right] d^2 +\mathcal{O}(Q_3), \nonumber\\
= Q_0-Q_1 d -\frac{1}{2} Q_2 d^2+\mathcal{O}(Q_3).
\label{eq05}
\end{multline}
where 
\begin{equation}
Q_j = \sum_{m=3} \frac{K_m}{(m-1)^j}   
\end{equation} 
for $j\ge1$. As $d$ approaches $1$, the term $K_2 \log(1-d)$ goes to infinity, which renders any error in the Taylor expansion of $Q(d)$ irrelevant. This makes the approximations above useable even if we ignore $\mathcal{O}(d^2)$ terms.

Putting all of the approximations above together, the ensuing approximate logarithm of the evidence $\mathcal{L}(\mathbf{n}|\alpha , d)$ is
\begin{multline}
\mathcal{L}(\mathbf{n}|\alpha , d) \approx  (K_1-1)\log d+\log\Gamma\left(\frac{\alpha}{d}+K_1\right)\\-\log\Gamma\left(\frac{\alpha}{d}+1\right) 
+\log\Gamma(1+\alpha)-\log\Gamma(N+\alpha)
\\+K_2 \log(1-d)-Q_1 d +\mathcal{O}\left(d^2\sum_{m=3} \frac{K_m}{(m-1)^2}\right),
\label{eq07}
\end{multline}
up to an additive constant. This is Eq.~(\ref{eq02:L}) in the main text.

\subsection{Maximum likelihood Entropy in terms of coincidences}
\label{A3:approxS}
To relate the conditional entropy, Eq.~(\ref{PYPEntropy}), to the Maximum Likelihood entropy estimator $S_0$, we need to rewrite the latter in terms coincidences. Utilizing the identity Eq.~(\ref{identity}), we  write 
\begin{multline}
 N\left[S_0-\log N\right] =-\sum_i n_i \log n_i =\\ -\sum_{m=2} (K_m-K_{m+1}) m \log m \\
= -K_2 (2 \log 2) \\-\sum_{m=3} K_m \big[ m \log m -(m-1)\log(m-1) \big].
\label{a1e5}
\end{multline}
Rewriting the expression in brackets as
\begin{equation}
 m \log m -(m-1)\log(m-1)= 1+ \psi(m)+\mathcal{O}(m^{-2}).
\label{a1e6}
\end{equation}
and plugging this into Eq.~(\ref{a1e5}), we finally obtain,
\begin{multline}
 N\left[S_0-\log N\right] = -K_2 \log 4 -(N-K_1-K_2)-\\\sum_{m=3} K_m \psi(m) +\mathcal{O}\left(\sum_m K_m/m^2\right).
\label{a1e7}
\end{multline}

\subsection{Mean posterior entropy approximation for the Pitman-Yor Process}\label{A2:approxS}

Similar to Appendix ~\ref{A1:approxL}, here we approximate the posterior entropy, Eq.~\ref{PYPEntropy}, in the limit of small $d$. To simplify the notation, we use the shorthand $S=\langle S|\bm{n},\alpha , d\rangle$ in this Appendix. Rearranging Eq.~(\ref{PYPEntropy}), we obtain
\begin{multline}
\displaystyle(\alpha+N)\left[S-\psi(N+\alpha+1)\right]=\\-\alpha\,\psi(1-d)-K_1\,d\,\psi(1-d)-\sum_i (n_i-d)\psi(n_i+1-d).
\label{a1e2}
\end{multline}

We now again use Eq.~(\ref{identity}) and a Taylor expansion in small $d$ to rewrite the last term on the right hand side of Eq.~(\ref{a1e2}):
\begin{align}
&K_1\,d\,\psi(1-d)-\sum_i (n_i-d)\psi(n_i+1-d) \nonumber\\
&=K_1\,d\,\psi(1-d)-\sum_{m=1} (K_m-K_{m+1}) (m-d)\psi(m+1-d) \nonumber\\
&=-\sum_{m=1} K_m \left[ (m-d)\psi(m+1-d) -(m-1-d)\psi(m-d) \right] \nonumber\\
&=-\sum_{m=1} K_m \left[ 1+\psi(m-d) \right] \nonumber\\
&=-\sum_{m=1}
K_m-\sum_{m=1} K_m \psi(m-d) \nonumber\\
&= -N-K_1\psi(1-d)-K_2 \psi(2-d)-\sum_{m=3} K_m \psi(m-d).
\label{a1e3}
\end{align}
where we used $\psi(m+1-d)=\left(\psi(m-d)+\frac{1}{m-d}\right).$

\flushend
Since $m\ge3$, we can Taylor expand the sum in this last term around $d=0$ to obtain
\begin{multline}
 \sum_{m=3} K_m \psi(m-d) \approx \sum_{m=3} K_m \psi(m)+d\,\sum_{m=3} K_m \psi'(m)\\+\mathcal{O}(d^2\sum_m K_m\psi''(m)).
\label{a1e4}
\end{multline}

Now  using the relations $\psi'(m) = \frac{1}{m-1}+\mathcal{O}(m^{-2})$ and the expression for $\sum_{m=3}K_m\psi(m)$ in Eq.~(\ref{a1e7}), we rewrite Eq.~(\ref{a1e4}) as 
\begin{multline}
\sum_{m=3} K_m \psi(m-d) \\\approx  K_2 \log 4 +(N-K_1-K_2)-N\left[S_0-\log N\right] \\
+d\,\sum_{m=3} \frac{K_m}{m-1}+\mathcal{O}(d^2,\sum_{m=3}K_m/m^2),
\end{multline}
where $\mathcal{O}(d^2,\sum_{m=3}K_m/m^2)$ means that we kept terms that are at most linear in $d$ and whose summands are at most proportional to $\sum_{m=3}K_m/m$. Plugging these approximation in Eq.~(\ref{a1e3}) and noticing that $Q_1 = \sum_{m=3} \frac{K_m}{m-1}$, we obtain
\begin{multline}
(\alpha+N)\left[S-\psi(N+\alpha+1)\right]
=\\ N(S_0-\log N) -\alpha\, \psi(1-d)
 +K_1 \left[-1-\,\psi(1-d) \right]\\+K_2 \left[-1-\psi(2-d)+\log 4 \right]
 -Q_1 \,d+ \mathcal{O}(d^2,\sum_{m=3}K_m/m^2),
\label{a1e10}
\end{multline}
which after isolating $S$ becomes  Eq.~(\ref{eq03:S}) of the main text.

\end{document}